\documentclass[preprint,showpacs,preprintnumbers,amsmath,amssymb]{revtex4}
                                                                                                                                                 
\usepackage{graphicx}
\usepackage{dcolumn}
\usepackage{bm}
\usepackage{ulem}
                                                                                                                                                
\begin{document}
                                                                                                                                               
                                                                                                                                               
\title{Tunneling Effects on Fine-Structure Splitting in Quantum Dot Molecules}
                                                                                                                                                 
\author{Hanz Y. Ram\'{\i}rez and Shun-Jen Cheng}
\affiliation{Department of Electrophysics, National Chiao Tung
  University, Hsinchu 300, Taiwan, Republic of China}
                                                                                                                                               
                                                                                                                                               
\date{\today}

\begin{abstract}
We theoretically study the effects of bias-controlled interdot tunneling in vertically coupled  quantum dots on the emission properties of spin excitons in various bias-controlled tunneling regimes. As a main result, for strongly coupled dots we predict substantial reduction of optical fine structure splitting without any drop in the optical oscillator strength. This special reduction diminishes the distinguibility of polarized decay paths in cascade emission processes suggesting the use of stacked quantum dot molecules as entangled photon-pair sources.
  
\end{abstract}
                                                                                                                      
\maketitle


Tunneling is a remarkable quantum property of microscopic particles that has no classical counterpart, which allows coupling between two objects spatially separated by a finite potential barrier. 
Currently, extending the analogy between atoms and 0D solid state systems, coupled quantum dots (QDs) are widely studied as artificial molecules where important properties of single dots are improved for optimization and scalability of applications. 
Recent examples of interesting and useful tunnel effects in coupled dot systems include the tunability of fluctuations in Kondo currents \cite{currents}, 
reduction of electronic spin decoherence by interaction with nuclear spin \cite{nuclear}, conditional dynamics of transitions \cite{newscience} and bias control of {\it g} tensors \cite{gtensor}.    
      
Currently, a highly desirable feature of QD-based photon emitters is the reduced fine structure splitting (FSS)  between the intermediate one-exciton (X) spin states. 
The FSS is widely believed to be a consequence of the electron-hole ({\it e-h}) exchange interaction caused by the intrinsic lack of perfect symmetry of QD structures \cite{review}. 
The FSSs make the two possible decay paths in bi-exciton cascade processes energetically distinguishable, and have become a main obstacle in the production of polarization-entangled photon pairs from QDs \cite{problem,entang1,zunger2,uk3}.   
Researchers have recently demonstrated significant reductions in the FSSs of single QDs using strain and post-annealing techniques, and the application of electric and magnetic fields  \cite{anealing,uk,lastpaul}. Nevertheless, to solve the ``which-path'' problem, the FSSs of bright X's (typically only $10^{1} \sim 10^{2} \mu$ eV) must be within the intrinsic broadenings of their emission lines (typically only $10^{0} \sim 10^{1} \mu$ eV) \cite{simultaneous}. In most experiments, however, it is not clear if the reduction of FSS is caused by the undoing of symmetry breaking or the reduction of e-h wave function overlap. The latter effect reduces not only the FSS but also the oscillator strength of e-h recombination, yielding narrow intrinsic broadening in the corresponding emission lines and actually inhibiting the generation of entangled photon pairs \cite{simultaneous,electric2}.

In this letter, we theoretically examine the effects of quantum tunneling in vertical QD molecules on the optical fine structure properties using the configuration interaction (CI) method. This study is based on a developed 3D model for coupled QDs that considers the both of mesoscopic (envelope function) and microscopic (Bloch function) nature of electrons and holes.   
As a result of quasi-resonant tunneling in stacked double dot systems, FSSs and photoluminescence (PL) intensities can by tuned by applying external bias fields and/or varying inter-dot distances. Remarkably, we predict a significant reduction of the optical FSSs in strongly coupled DQDs with small inter-dot distances without any decrease in the optical oscillator strength.
 
\begin{figure}
\begin{center}
\scalebox{2}{\includegraphics{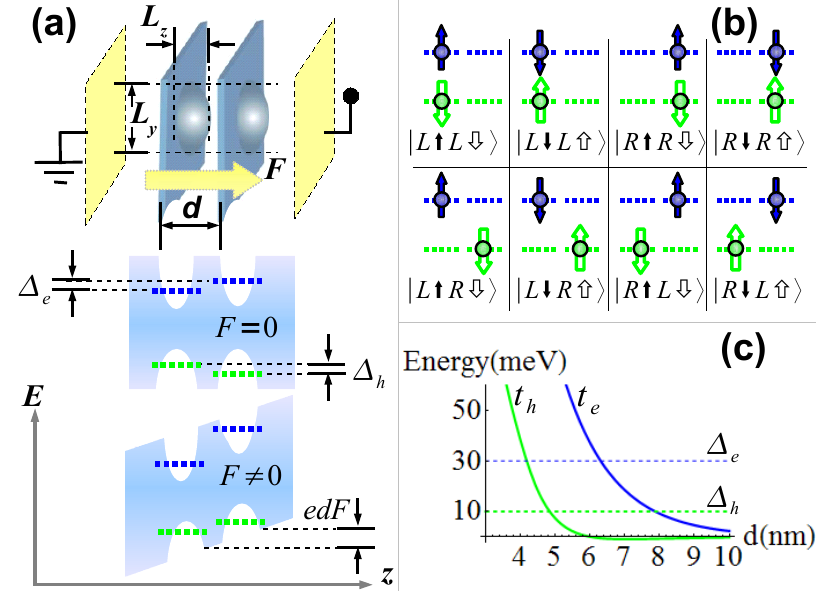}}
\caption{(Color online) Schematic diagrams of (a) a double QD structure and (b) spin exciton configurations. (c) The calculated hopping parameters, $t_{e}$ blue (dark) and $t_{h}$ green (light), vs. interdot distance $d$. Horizontal dashed lines: the values of $\Delta_{e}$ and $\Delta{h}$ considered throughout this work.}
\label{fig1} 
\end{center}
\end{figure}

Let us consider a pair of vertically stacked quantum dots along the growth $z-$axis, separated by an inter-dot distance $d$ and subject to an applied electric field $F$, as shown in Fig. \ref{fig1}(a) \cite{chang}. The {\it e-h} Hamiltonian for a single spin exciton in a coupled double QD is written as

\begin{center}
\begin{eqnarray}
\label{totham}
H&=&\sum_{j,\sigma}(\varepsilon_{j}^{e}+eFz_j)c_{j\sigma}^\dag c_{j\sigma}+\sum_{n,\chi}(\varepsilon_{n}^{h}-eFz_n)h_{n\chi}^\dag h_{n\chi}\nonumber \\
&-&\sum_{j\in L,k\in R,\sigma}t_{jk}^e(c_{j\sigma}^\dag c_{k\sigma} + c_{k\sigma}^\dag c_{j\sigma}) \nonumber \\
&-& \sum_{n\in L,m\in R,\sigma}t_{nm}^h(h_{n\chi}^\dag h_{m\chi} + h_{m\chi}^\dag h_{n\chi}) \nonumber \\
&-& \sum_{kmnj,\sigma\chi}V_{kmnj}^{eh}c_{j\sigma}^\dag h_{m\chi}^\dag h_{n\chi}c_{j\sigma} \nonumber \\
&-& \sum_{kmnj,\sigma \chi \chi' \sigma'}V_{k\sigma,m\chi,n\chi',j\sigma'}^{eh,ex}c_{k\sigma}^\dag h_{m\chi}^\dag h_{n\chi'}c_{j\sigma'} \, , 
\end{eqnarray}
\end{center}

where the composite indexes $j,k$ ($n,m$) denote the electron (valence hole) orbitals and dot positions ($L$/$R$ for the left/right dot), $\sigma=\uparrow /\downarrow$ ($\chi=\Uparrow/ \Downarrow$) represents electron (hole) spin with $s_z=\frac{1}{2}/-\frac{1}{2}$ ($j_z=\frac{3}{2}/-\frac{3}{2}$), $c_{j\sigma}^{\dag}$  and $c_{j\sigma}$ ($h_{n\chi}^{\dag}$ and $h_{n\chi}$) are the electron (hole) creation and annihilation operators respectively, $\varepsilon_{i}^{e}$ ($\varepsilon_{n}^{h}$) is the kinetic energy of an electron (a valence hole), $e$ is the unit charge, and $z_{j\in L}=0$ 
$(z_{j\in R}=d)$ is the $z-$position of the left (right) dot. Here, the valence hole orbitals of the highly quantized strained dots are assumed to be purely heavy-hole like. The terms with the hopping parameters ($t_{jk}^e$, $t_{nm}^h$) describe the (spin-conserved) carrier tunneling between adjacent dots. The matrix elements of {\it conventional} {\it e-h} Coulomb interaction and the {\it e-h exchange} interactions are $V_{kmnj}^{eh} \equiv  \int \int d^3{r_{1}} d^3{r_{2}} \Phi_{k}^{e\ast}(\vec{r_{1}}) \Phi_{m}^{h\ast}(\vec{r_{2}}) \frac{e^{2}}{4 \pi \epsilon r_{12}}  \Phi_{n}^{h}(\vec{r_{2}}) \Phi_{j}^{e}(\vec{r_{1}})$ and
$V_{k\sigma,m\chi,n\chi',j\sigma'}^{eh-ex}  \equiv \int  \int d^3{r_{1}} d^3{r_{2}} \Phi_{k}^{e\ast}(\vec{r_{1}}) u_{c \sigma }^\ast(\vec{r_{1}}) 
\Phi_{m}^{h\ast}(\vec{r_{2}}) u_{v \chi}^\ast(\vec{r_{2}})$ $\times$ $
 \frac{e^{2}}{4 \pi \epsilon \,\, r_{12}}  \Phi_{n}^{h}(\vec{r_{1}}) u_{v \chi'}(\vec{r_{1}}) \Phi_{j}^{e}(\vec{r_{2}}) u_{c \sigma'}(\vec{r_{2}})$, respectively, where $\Phi_\alpha$ are single-particle envelope wave functions,  $u_{c \sigma}$ ($u_{v \chi}$) are the electron (hole) Bloch functions, $\epsilon$ is the dielectric constant and $ r_{12} \equiv \mid \vec{r_{1}} - \vec{r_{2}} \mid$.
Remarkably, after undergoing an {\it e-h} exchange interaction, an electron or a hole could lose its spin conservation.
Within the dipole-dipole approximation, the long-range part of the {\it e-h} exchange interaction is given by $V_{kmnj}^{eh-ex (Lr)} \equiv \delta_{1}^{kmnj}  
\approx \frac{3 e^{2} \hbar^{2} E_{p}}{2 \epsilon m_{0} E_{g}^{2}} 
\int \int d^3\vec{r}_1 d^3\vec{r}_2  
\Phi_{k}^{e\ast}(\vec{r}_1) \Phi_{m}^{h\ast}(\vec{r}_2)  
\Phi_{n}^{h}(\vec{r}_1) \Phi_{j}^{e}(\vec{r_2}) 
 [(y_1-y_2)^{2} 
- (x_1-x_2)^2 + 2i(x_1-x_2)(y_1-y_2)]/(r_{12})^5$ ,

where $E_p$ is the conduction-valence band interaction energy, $E_g$  the band gap energy, and $m_0$ the mass of a free electron \cite{poem}.

Based on the lowest single-particle orbitals of single dots, eight spin-X configurations are constructed, as displayed in Fig. \ref{fig1}(b).
To analyze further the (linear) polarization of emitted light, a new basis is defined by the linear transformation of the configurations according to the parity symmetry: $|LL\pm\rangle \equiv \frac{1}{\sqrt{2}}(|L\uparrow L\Downarrow\rangle \pm |L\downarrow L\Uparrow\rangle)$, $|RR\pm\rangle \equiv \frac{1}{\sqrt{2}}(|R\uparrow R\Downarrow\rangle \pm |R\downarrow R\Uparrow\rangle)$,$|LR\pm\rangle \equiv \frac{1}{\sqrt{2}}(|L\uparrow R\Downarrow\rangle \pm |L\downarrow R\Uparrow\rangle)$,$|RL\pm\rangle \equiv \frac{1}{\sqrt{2}}(|R\uparrow L\Downarrow\rangle \pm |R\downarrow L\Uparrow\rangle)$.  In the redefined basis, the $8 \times 8$ Hamiltonian can be decomposed into two decoupled $4 \times 4$ matrices. Only the configurations with positive (negative) parity are associated with a $\pi_x$- ($\pi_y$-) polarized light emission. In the basis ordered by $|LL\pm\rangle$, $|RR\pm\rangle$, $|LR\pm\rangle$, $|RL\pm\rangle$, the decoupled $4 \times 4$ Hamiltonian matrix is 

{\tiny
\begin{equation}
\label{hammat}
\hat{H_{\pm}} = \left(
\begin{array}{cccc}
-V_{eh} \mp \delta_{DD} & \mp \delta_{II} & -t_h  & -t_e  \\
\mp \delta_{II} & -V_{eh}+\Delta_e+\Delta_h \mp \delta_{DD} & -t_e  & -t_h  \\
-t_h & -t_e  & -eFd + \Delta_{h} \mp \delta_{II} & \mp \delta_{II}\\ 
-t_e &  -t_h  & \mp\delta_{II} & eFd  + \Delta_{e} \mp \delta_{II}
\end{array}
\right) \hspace*{1ex} ,
\end{equation}}

where the kinetic energy offset $\varepsilon_{L}^{e/h}+\varepsilon_{L}^{e/h}$ is removed for brevity, $\Delta_{e/h}\equiv \varepsilon_{R}^{e/h}- \varepsilon_{L}^{e/h}$ denotes the difference between kinetic energies of  the two adjacent dots due to the inevitable slight differences in size, shape or chemical composition,  $V_{eh}\equiv V_{LLLL}^{eh}=V_{RRRR}^{eh}$ denotes the direct Coulomb interaction between an {\it e-h} pair in the same single dot, and $\delta_{DD}\equiv \delta_1^{RRRR}=\delta_1^{LLLL}$ ($\delta_{II}\equiv \delta_1^{LRRL}=\delta_1^{RLLR}\approx\delta_1^{LLRR}=\delta_1^{RRLL}$) is the long range {\it e-h} exchange interaction in a direct X (an indirect X). 
Previous studies concerning {\it e-h} exchange matrix elements in single and laterally coupled dots use 2D approaches \cite{poem,ivchenko,shangai}. 
However, a fully three-dimensional formulation, including dot height and interdot distance, is required to accurately consider tunneling effects in stacked QD molecules.
Within the 3D parabolic model for the confining potentials of single QDs, the single-particle wave functions of the lowest orbitals of single dots can be  described by $\Phi_{L/R}(x,y,z)= (\pi^{\frac{3}{2}}l_{x}l_{y}l_{z})^{-1/2} \exp[-\frac{1}{2}((\frac{x}{l_{x}})^{2}+(\frac{y}{l_{y}})^{2}+(\frac{z-z_{L/R}}{l_{z}})^{2})]$, characterized by the wave function extents $l_{\alpha=x,y,z}$. Accordingly, we derive  $V_{eh}\approx \frac{e^2}{4\pi\epsilon} \frac{1}{l}\frac{\sqrt{2}\sin^{-1}(1-a^2)}{\sqrt{\pi}(1-a^2)}$, 
\newline $\delta_{DD}=\frac{e^{2} \hbar^{2} E_{p}}{2 \sqrt{2 \pi} \epsilon m_{0} E_{g}^{2}} \frac{(l_{x}-l_{y})}{l_{y}} \frac{1}{l_{y}^{2} l_{z}} \, , $
and
$\delta_{II}=\delta_{DD} \hspace*{1ex} e \hspace*{1ex} ^{-\frac{d^{2}}{2l_{z}^{2}}} \, , $ 
for a slightly deformed DQD ($\xi \equiv \frac{l_x-l_y}{l_y} \ll 1 \neq 0$ , where $l\equiv (l_x+l_y)/2, a\equiv l_z/l$). The values of  $t_e$ and $t_h$, are determined by the formulation presented in Refs. \cite{observation} and \cite{nacho}, respectively. 
Figure \ref{fig1}(c) shows the calculated $t_{e}$ and the $t_{h}$ as functions of $d$ \cite{doty}. 

\begin{figure} [t]
\scalebox{2}{\includegraphics{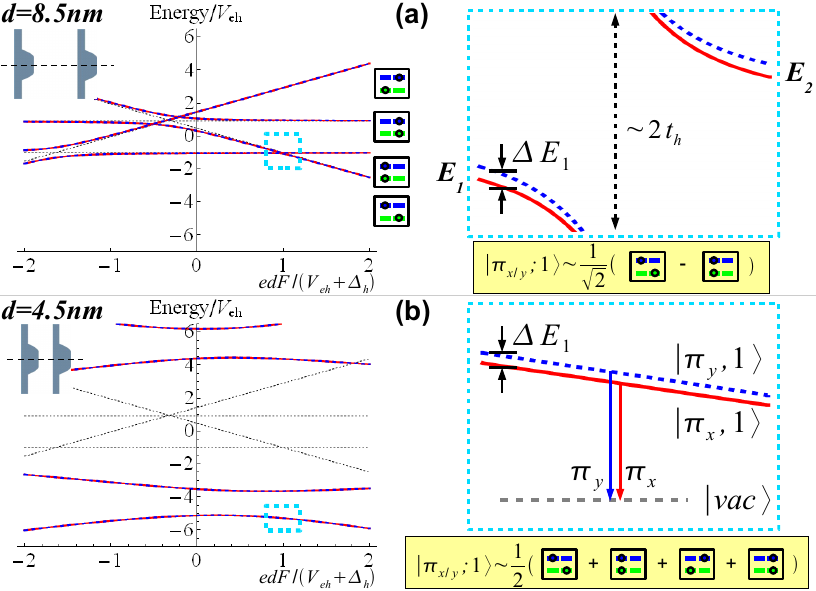}}
\caption{(Color online) Calculated energy spectra vs. bias field $F$ of (a) a weakly coupled DQD with $d=8.5$nm and (b) a strongly coupled DQD with $d=4.5$nm. Straight dashed lines describe the energy spectrum of a decoupled DQD. Insets: the magnified fine structures of the energy spectra of the DQDs at near resonance and the (schematic) configuration intermixings of the lowest exciton states.}
\label{fig2} 
\end{figure}

The energy spectrum $\{E_{\pi_x,i}\}$ ($\{E_{\pi_y,i}\}$) of the exciton states $|\pi_x;i\rangle$ ($|\pi_y;i\rangle$) for the $\pi_x$ ($\pi_y$)-polarized light emission is calculated by diagonalizing $H_{+}$ ($H_{-}$) in Eq.(\ref{hammat}) \cite{parameters}. In the combined energy spectrum, $\{E_{\pi_x,i}, E_{\pi_x,i}\}$,  each level is a doublet of the spin X states, $|\pi_x;i\rangle$ and $|\pi_y;i\rangle$, which are split by an FSS $\Delta E_i \equiv E_{\pi_y,i} -E_{\pi_x,i}$ [inset of Fig. \ref{fig2}(a)]. The $\pi_x$($\pi_y$)-linear-polarized photoluminescence (PL) spectra are obtained using Fermi's golden rule: $I_{x(y)}(\omega) =  \sum_{i} F(E_{i},T) |\langle 0 | P^{-}_{x(y)} | \pi_x(\pi_y); i \rangle |^{2} \delta(E_{\pi_x(\pi_y),i}-\hbar \omega)$, where the subscript $i$ ($f$) denotes initial (final) states of the PL transition, $\omega$ is the frequency of the emitted photon, the operator 
$P_{x}^{(-)}=\sum_{n,j} S_{n,j} (h_{n\Uparrow} c_{j\downarrow} + h_{n\Downarrow} c_{j\uparrow})$ [$P_{y}^{(-)}=-i\sum_{n,j} S_{n,j} (h_{n\Uparrow} c_{j\downarrow} - h_{n\Downarrow} c_{j\uparrow})$]
describes the all possible {\it e-h} recombinations that produce the $\pi_x [\pi_y]$ linear polarized PL,  $S_{n,j}=\int d^3r \Phi_{n}^{h\ast}(\vec{r}) \Phi_{j}^{e}(\vec{r})$ is the e- and h-wave function overlap, and $F(E_{i},T)=\exp(-E_i/k_BT)/[\sum_{l}\exp(-E_l/k_BT)]$ is the probability of occupation of state $|i\rangle$, where $k_B$ is the Boltzmann constant and $T$ is temperature.

\begin{figure} [b]
\scalebox{2}{\includegraphics{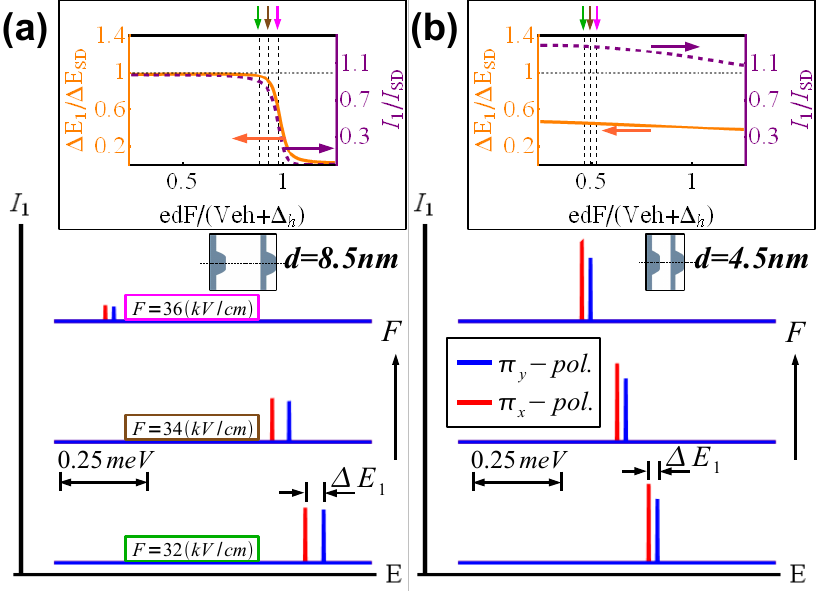}}
\caption{(Color online) (a) Calculated polarized PL spectra of a weakly coupled DQD with $d=8.5$nm under the electric biases $F$ at near resonance at $T=10K$. The inset: the normalized FSS $\Delta E_1/\Delta E_{SD}$ (orange solid line) and intensity $I_1/I_{SD}$ (purple dashed line) of the main PL spectral lines as functions of $F$, where $\Delta E_{SD}$ and $I_{SD}$ denotes the FSS and intensity of the main PL line of a single dot. The considered biases in the calculated PL spectra are indicated with vertical arrows in the inset.  (b) The calculated results same as (a) but for a strongly coupled DQD with $d=4.5$nm.}
\label{fig3} 
\end{figure}

Figure \ref{fig2}(a) shows the calculated energy spectra of a weakly coupled DQD with $d=8.5$nm at various applied biases.  The corresponding hopping parameters are $t_e=6.4$meV and $t_h=-0.6$meV. Under the weak coupling (WC) condition ($\Delta_e \gg t_e > t_h$), the $4\times 4$ Hamiltonian matrix, Eq.(\ref{hammat}), can be decomposed into two $2\times 2$ blocks that are coupled only by a relatively weak electron hopping ($t_e/\Delta_e \ll 1$). 
Thus, the Hamiltonian matrix for the two lowest spin-X states can be approximated as the following $2\times 2$ block:  

\begin{equation}
\label{hammat1}
{\scriptsize \hat{H}^{WC}_{\pm}} = \left(
\begin{array}{cc}
-V_{eh} \mp \delta_{DD} & -t_h \\
-t_h & -eFd + \Delta_h \mp \delta_{DD}
\end{array}
\right) 
\end{equation}

with respect to the basis $|LL\pm\rangle$ and $|LR\pm\rangle$. Equation (\ref{hammat1}) is actually equivalent to the widely used solvable three-orbital model for DQDs \cite{science}. The eigen states of Eq.(\ref{hammat1}) are hybridized by the optically active X-configuration $|LL\pm\rangle$ and the inactive configuration $|LR\pm\rangle$, determined by the bias-controlled detuning from resonance ($|edF-(\Delta_h+V_{eh})|$). Expanding the X eigen states in the used basis for Eq.(\ref{hammat}), i.e. $|\pi_x;i\rangle =\sum_{nj} C^{x}_{nj,i} |nj+\rangle$ and $|\pi_y;i\rangle =\sum_{nj} C^{y}_{nj,i} |nj-\rangle$, the intensities and the FSS associated with the lowest spectral lines are given by $I_{1}\approx F(E_{1},T) (C_{LL,1} S_{D} + C_{LR,1} S_{I})^{2}$ and $\Delta E_{1}\approx 2 (C_{LL,1}^{2} \delta_{DD} + C_{LR,1}^{2} \delta_{II})$, where $C_{LL,1}\equiv C_{LL,1}^{x}=C_{LL,1}^{y}$ ($C_{LR,1}\equiv C_{LR,1}^{x}=C_{LR,1}^{y}$) are the expansion coefficients associated with the bright (dark) X configurations $|LL\pm\rangle$ ($|LR\pm\rangle$) and $S_D\equiv S_{LL}=S_{RR}\approx 1$ ($S_I\equiv S_{LR}=S_{RL}=e \hspace*{1ex} ^{-\frac{d^{2}}{4 l_{z}^{2}}}$) is the {\it e-h} wave function overlap in a direct-X (an indirect-X) configuration.  Accordingly, both of the $I_1$ and the $\Delta E_1$ of a weakly coupled DQD ($S_I \ll S_D$ and $\delta_{II} \ll \delta_{DD}$) are mainly proportional to $C_{LL,1}^2$ and should depend similarly on applied bias fields. Figure \ref{fig3}(a) shows the calculated polarized PL spectra of the weakly coupled DQD at some bias fields in the near-resonance regime and the inset shows the $F$-dependences of the $I_1$ and $\Delta E_1$.

At very low bias ( $| edF/ (V_{eh}+\Delta_h)| \ll 1$), the ground states of the exciton are $|\pi_x;1\rangle \approx |LL+\rangle $ and $|\pi_y;1\rangle \approx |LL-\rangle$. The intensity (FSS) of the corresponding linear polarized emission lines is $I_1\approx (S_D)^2$ ($\Delta E_1 \approx 2\delta_{DD}$), approaching the value of the intensity (FSS) of the lowest spectral lines of a single dot,  $I_{SD}$ ($\Delta E_{SD}$). 
At near resonance ($ edF/(V_{eh}+\Delta_h) \approx 1$), where $|\pi_x;1\rangle \approx \frac{1}{\sqrt{2}}(|LL+\rangle - |LR+\rangle)$ and $|\pi_y;1\rangle \approx \frac{1}{\sqrt{2}}(|LL-\rangle - |LR-\rangle)$, only the hole in the exciton can be transferred between dots while the electron is stably localized in the left dot. 
The intensity (FSS) of the corresponding polarized emission lines is $I_1\approx (S_D + S_I)^2/2$ ($\Delta E_1 \approx \delta_{DD}+\delta_{II}$), which is only about $50\%$ of that for a single dot. 
The resonant inter-dot tunneling of a single hole significantly reduces the overlap of the electron and hole wave functions, leading to not only the decrease in the optical FSS but also the oscillator strength of an {\it e-h} recombination. 
The decreased oscillator strength of {\it e-h} recombination reduces the intrinsic broadening width of the main X lines. Such an FSS reduction however does not support the feasibility of the dot-based entangled photon pair source devices \cite{new,uk2}.

Figure \ref{fig2}(b) shows the energy spectra  of a strongly coupled DQD with small distance $d=4.5$nm. The corresponding hopping parameters are $t_e=106$meV and $t_h=18.2$meV. Figure \ref{fig3}(b) plots the normalized $I_1$ and $\Delta E_1$ of the lowest spectral lines vs. $F$.  Generally, the strongly coupled DQD have smaller FSS $\Delta E_1$ but larger $I_1$ than single dots or weakly coupled dot molecules. 
 
In the strong coupling (SC) limit ($t_{\beta}\gg\Delta_{\beta}$), both electrons and holes can be transferred between dots over a very wide range of detuning and 
the Hamiltonian, Eq.(\ref{hammat}), can be approximately written as
 
\begin{equation}
\label{higcou}
\hat{H}_{\pm}^{SC} \approx \left(
\begin{array}{cccc}
0 & 0 & -t_{h} & -t_{e}\\
0 & 0 & -t_{e} & -t_{h}\\ 
-t_{h} & -t_{e} & 0 & 0 \\
-t_{e} & -t_{h} & 0 & 0 \\
\end{array}
\right) \hspace*{1ex}.
\end{equation}

The lowest eigenstates for Eq.(\ref{higcou}) are $\frac{1}{2}(|LL\pm\rangle+|RR\pm\rangle+|LR\pm\rangle+|RL\pm\rangle )$, highly intermixing all X-configurations. Accordingly, we have $I_1\approx (S_D + S_I)^2$  and $\Delta E_1 \approx \delta_{DD}+2\delta_{II}$, {\it i.e.} that the FSS is only about one half of the magnitude of $\Delta E_{SD}$ but the intensity of the polarized emission lines is slightly larger than $I_{SD}$. 
In the strong coupling regime, not only valence holes but also electrons are spread over the two coupled dots. The simultaneous {\it e-h} resonance transfers between dots enlarges the optically active volume and increase the mean distance $\langle r_{12}\rangle$ in the long ranged {\it e-h} exchange interactions, resulting in the larger $I_1$ and smaller $\Delta E_1$.

\begin{figure}
\scalebox{2}{\includegraphics{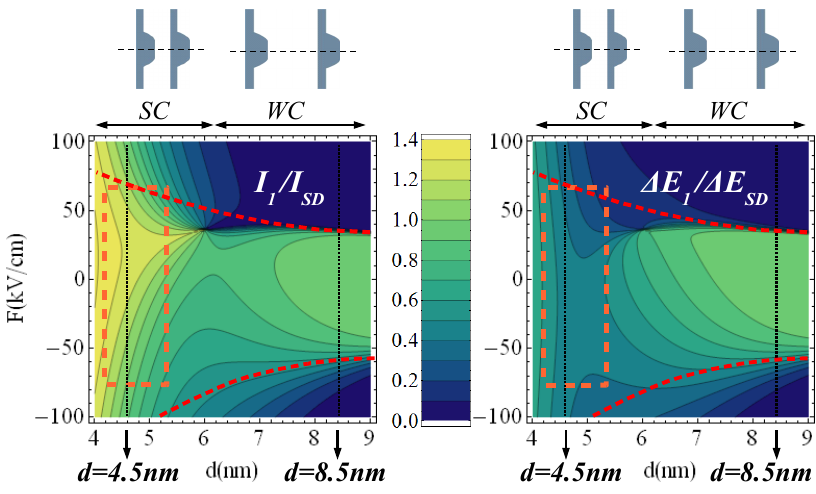}}
  \caption{(Color online) Normalized intensity $I_{1}/I_{SD}$ (left) and FSS $\Delta E_{1}/\Delta E_{SD}$ (right) of the lowest PL spectral lines of coupled DQDs, as functions of inter-dot distance $d$ and bias field $F$. The dashed line boxes highlight the reduced $\Delta E_1$ and increased $I_1$ of the strongly coupled DQDs. The vertical dotted lines indicate  $d=4.5$nm and $d=8.5$nm for which Figs.~\ref{fig2} and \ref{fig3} are calculated. The red dashed line in the upper (lower) half plane indicates the hole (electron) resonances. }
\label{fig4} 
\end{figure}

Figure \ref{fig4} plots the normalized $I_1$ and $\Delta E_1$ (by $I_{SD}$ and $\Delta E_{SD}$) of DQDs as functions of the inter-dot distance $d$ and applied bias fields $F$.  
In the WC regime, as discussed previously, $I_1$ and $\Delta E_1$ depend similarly on $F$. As a DQD is driven into the SC regime, $I_1$ are markedly increased and the FSS is reduced to only $\sim 50\%$ of $\Delta E_{SD}$ (see the regions  highlighted by in dash-line boxes) \cite{chang}. The increased $I_1$ and reduced $\Delta E_{1}$ are robust against the detuning, being almost insensitive to $F$. The lower part of Fig.~\ref{fig4} plots the results obtained for DQDs at negative $F$, which drives electron inter-dot transfers. The results for the DQDs at negative $F$ show similar physical features to those at positive $F$. The only slight difference is that the near resonance region is wider than that for the DQDs at positive $F$ because of the larger magnitude of tunneling coupling for electrons \cite{scheibner,peteers}.  
 
In summary, this study discusses the effects of tunnel coupling on photon emission from spin excitons in vertically stacked double quantum dots. Results show that an increase in the optically active volume and electric charge deconcentration caused by simultaneous electron and hole transfers between dots significantly inhibits the optical fine structure splitting of coupled QDs in the strong coupling regime without any decrease in optical oscillation strength. This tunneling-driven FSS reduction is robust against the bias-controlled detuning from resonance, making strongly coupled vertical quantum dot molecules  better cascade decay sources of entangled photon pairs than single dots.

The authors would like to thank the National Science Council of Taiwan for
financially supporting this research under Contract No. NSC-95-2112-M-009-033-MY3. Wen-Hao Chang (NCTU) is appreciated for his valuable discussions.

\end{document}